# Multi-GPU implementation of a time-explicit finite volume solver for the Shallow-Water Equations using CUDA and a CUDA-Aware version of OpenMPI

Vincent Delmas[a,b], Azzedine Soulaïmani[a,*]

[a]*Department of Mechanical Engineering, École de Technologie Supérieure (ÉTS), 1100 Notre-Dame Ouest, Montréal, QC H3C 1K3, CANADA*
[b]*Département de Mathématique et Mécanique, École nationale supérieure d'électronique, informatique, télécommunications, mathématique et mécanique de Bordeaux (ENSEIRB-MATMECA), 1 avenue du Dr. Albert Schweitzer, 33402 Talence Cedex, FRANCE*

## ARTICLE INFO

## Abstract

This paper shows the development of a multi-GPU version of a time-explicit finite volume solver for the Shallow-Water Equations (SWE) on a multi-GPU architecture. MPI is combined with CUDA-Fortran in order to use as many GPUs as needed. The METIS library is leveraged to perform a domain decomposition on the 2D unstructured triangular meshes of interest. A CUDA-Aware OpenMPI version is adopted to speed up the messages between the MPI processes. A study of both speed-up and efficiency is conducted; first, for a classic dam-break flow in a canal, and then for two real domains with complex bathymetries: the Mille Îles river and the Montreal archipelago. In both cases, meshes with up to 13 million cells are used. Using 24 to 28 GPUs on these meshes leads to an efficiency of 80% and more. Finally, the multi-GPU version is compared to the pure MPI multi-CPU version, and it is concluded that in this particular case, about 100 CPU cores would be needed to achieve the same performance as one GPU.

## 1. Introduction

Floods are among the costliest natural disasters. Whether due to tsunamis, ruptured dams, or heavy rainfall, they affect large areas and often require the evacuation of many people. Governments and agencies must therefore develop reliable and accurate maps of flood risk areas as part of their preventive measures (Raja and Elshorbagy, 2018; Das and Umamahesh, 2018; Haltas et al., 2016; Tsai and Yeh, 2017). Hence, predictive simulations should quantify the uncertainties that may arise from multiple sources (such as the boundary conditions, the geometry, the physical parameters, etc.) and that propagate through the modeling system. Since the uncertainties propagation methods usually require a large dataset of high-fidelity solutions, these should be computed in a reasonable time-frame.

Several numerical simulation codes have been developed to model floods, usually by solving the Shallow-Water Equations (Soulaimani et al., 2002; Toro, 2001; Audusse et al., 2004a; Bradford and Sanders, 2002; Brufau et al., 2004; Zokagoa and Soulaïmani, 2010; Loukili and Soulaimani, 2007). These codes need to be capable of computing solutions for large domains very quickly. Rapid computations are especially required in the context of uncertainty propagation or inverse analysis studies, and more generally for the case of an extreme emergency event. Parallel computing is deemed essential to achieve this goal, often using GPUs to further speed up the computations.

In De la Asunción et al. (2010); De la Asunción et al. (2013); Niksiar et al. (2014); Ayyad et al. (2020), a single GPU was used to speed up the computations. Speed-ups on the order of 10 to 40 times were achieved by the GPU versions compared to their CPU sequential counterparts. In Vacondio et al. (2014), with the use of better GPUs, even greater speed-ups are achieved. One GPU is used to solve the SWE using CUDA C, C++ or Fortran in Brodtkorb et al. (2012, 2010); Escalante et al. (2018) and using OpenCL in Smith and Liang (2013).

The need for faster computations and larger domains led to the use of multiple GPUs per simulation. In order to meet these expectations, the use of GPU programming languages like CUDA C, C++, Fortran or OpenCL was combined with some CPU-level parallelism, such as OpenMP or MPI. The most popular approach is to use MPI coupled with CUDA C, C++ or Fortan and domain decomposition. In this approach, each MPI process solves the problem on a sub-domain using the GPU it is associated with. Such approaches can be found in Komatitsch et al. (2010); Jacobsen et al.; Jacobsen and Senocak (2011); Lai et al. (2019); Viñas et al. (2013); Turchetto et al. (2020).

---

*Corresponding author

This paper describes a methodology for porting a finite volume solver for the SWE on a multi-GPU architecture. The methodology is quite general and thus applicable for other solvers, especially with explicit time discretization. We use this methodology to port our in-house code **CuteFlow** (Loukili and Soulaïmani, 2007; Zokagoa and Soulaïmani, 2010; Suthar and Soulaimani, 2018; Jacquier et al., 2021) for the resolution of the SWE on a multi-GPU architecture using CUDA Fortran, a CUDA-Aware version of OpenMPI Gabriel et al. (2004) and METIS Karypis and Kumar (2009) to perform the domain decomposition.

The article is divided into several sections. In section 2, we present the SWE and their resolution using a time-explicit finite volume method. The Riemann solvers used during this step are presented in section 3, as well as the treatment of bathymetric and friction source terms. Next, in section 4 we tackle the heart of multi-GPU porting, starting with the pre-processing required for the domain decomposition. We use the METIS library (Karypis and Kumar, 2009) to performd the decomposition and devote particular attention to the numbering of the cells to be sent and received by each sub-domain. The use of OpenMPI (Gabriel et al., 2004), in combination with CUDA, is then presented in section 5. We show in section 5.3 the general functioning of our in-house code and how we overlapped the computations with the MPI memory exchange. The results obtained on different domains are presented in section 6. We begin with a classic dam-break case, followed by the domain of the Mille Îles river and then the domain of the archipelago of Montreal. We calculate the speed-up and efficiencies for different mesh sizes and comment on the scaling of our proposed multi-GPU version. The performance of the multi-GPU version is then compared to that of of the pure MPI multi-CPU version. A discussion of the usefulness of the multi-GPU version concludes section 6. We end with our conclusions and recommendations for further work in section 7.

## 2. The Shallow-Water Equations

The Shallow-Water Equations system is presented here. The convention for describing the bathymetry $b$, the surface $s$ and the water height $h$ is illustrated on Figure 1.

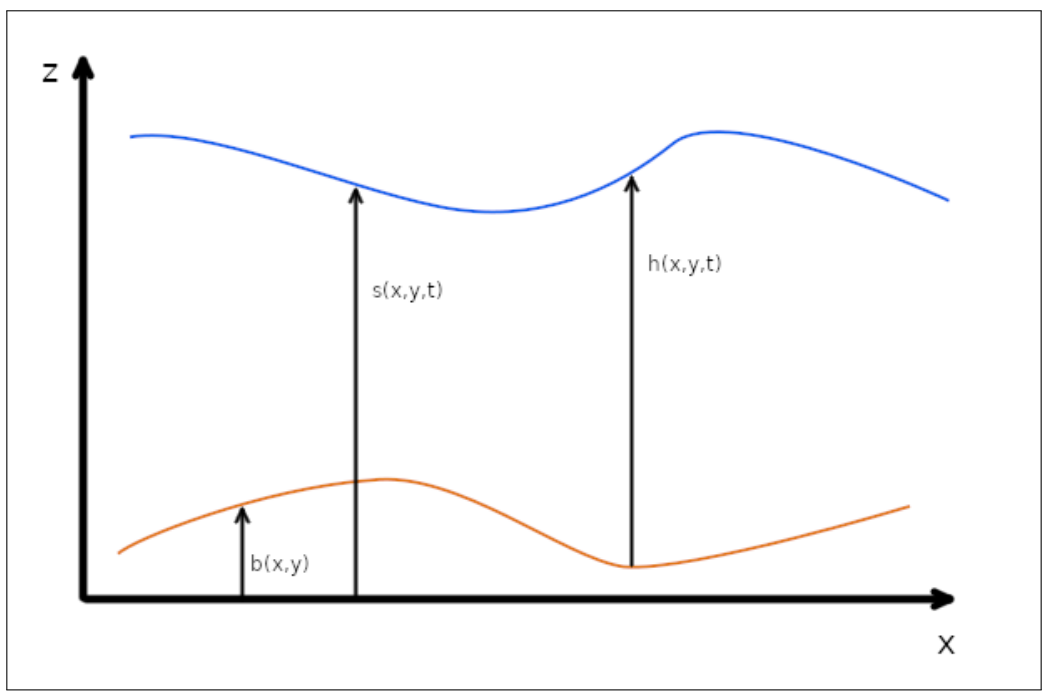

**Figure 1:** Illustration of the notations

The Shallow-Water Equations system is written as

$$U_t + F(U)_x + G(U)_y = S(U),$$ (1)

with

$$U = \begin{bmatrix} h \\ h\bar{u} \\ h\bar{v} \end{bmatrix}, \qquad F(U) = \begin{bmatrix} h\bar{u} \\ h\bar{u}^2 + \frac{1}{2}gh^2 \\ h\bar{u}\bar{v} \end{bmatrix},$$

$$G(U) = \begin{bmatrix} h\bar{v} \\ h\bar{u}\bar{v} \\ h\bar{v}^2 + \frac{1}{2}gh^2 \end{bmatrix}, \qquad S(U) = \begin{bmatrix} s_1 \\ s_2 \\ s_3 \end{bmatrix}.$$

where $\bar{u}$ and $\bar{v}$ are the depth averaged velocities on the x and y direction, $h$ is the height of the water column as defined in Figure 1, and g is the gravitational acceleration.

The system can be represented in integral form in order to accept the discontinuities, as in the following expression

$$\frac{\partial}{\partial t} \int_\Omega U \, dV + \int_{\partial\Omega} n.H(U) \, dS = \int_\Omega S(U) \, dV,$$ (2)

with $H(U) = (F(U), G(U))$.

The source term of (1) can be separated into two, one part for bathymetry and one for friction. We will use the following notations

$$S(U) = S_O(U) + S_f(U)$$ (3)

where the bathymetry term $S_O$ is

$$S_O = (0, -ghb_x, -ghb_y)$$ (4)

and the friction term $S_f$ is

$$S_f = (0, -gh\frac{m^2\bar{u}\sqrt{\bar{u}^2 + \bar{v}^2}}{h^{4/3}}, -gh\frac{m^2\bar{v}\sqrt{\bar{u}^2 + \bar{v}^2}}{h^{4/3}}).$$ (5)

in which $m$ is Manning's roughness coefficient.

## 3. Numerical resolution of the SWE by the finite volume method

This section presents the resolution of the SWE by the finite volume method (Loukili and Soulaimani (2007); Ata et al. (2013); Audusse and Bristeau (2005); Toro (2001)). We use a finite volume method centered on the cells, realized by combining the HLLC scheme and the semi-implicitation of the friction terms of Loukili and Soulaimani (2007) with the treatment of the bathymetry terms presented in Ata et al. (2013), an approach inspired by Audusse and Bristeau (2005).

### 3.1. Finite volume discretization

The finite volume method is based on a tiling of the study area in volumes; we chose to use triangular volumes. We start by integrating the SWE on each volume $\Omega_i$ which, by applying the divergence theorem, gives

$$\int_{\Omega_i} \frac{\partial U}{\partial t} \, dV + \int_{\partial\Omega_i} H(U).n_i \, dS = \int_{\Omega_i} S(U) \, dV$$ (6)

with $H(U) = (F, G)^T$, and $n_i$ is the unit normal of $\partial\Omega_i$ outwards of $\Omega_i$.

We then use the following definitions

$$U_i = \frac{1}{|\Omega_i|} \int_{\Omega_i} U \, dV,$$ (7)

and

$$S_i(U) = \frac{1}{|\Omega_i|} \int_{\Omega_i} S(U) \, dV, \tag{8}$$

which transform (6) into

$$|\Omega_i| \frac{dU_i}{dt} = -\sum_{j=1}^{3} L_{ij} H(U).n_{ij} + |\Omega_i| S_i(U), \tag{9}$$

with $|\Omega_i|$ the area of $\Omega_i$ (triangular), $L_{ij}$ the length of the side $j$ of $\Omega_i$ and $n_{ij}$ the unit normal of the side $j$ outwards of $\Omega_i$.

We can also separate the source term $S_i$ into two parts, $S_{O_i}$ for bathymetry and $S_{f_i}$ for friction

$$\begin{cases} S_{O_i}(U) = & \dfrac{1}{|\Omega_i|} \displaystyle\int_{\Omega_i} S_O(U) \, dV, \\[2mm] S_{f_i}(U) = & \dfrac{1}{|\Omega_i|} \displaystyle\int_{\Omega_i} S_f(U) \, dV. \end{cases} \tag{10}$$

Equation (9) can then be written as

$$|\Omega_i| \frac{dU}{dt} = -\sum_{j=1}^{3} L_{ij} H(U).n_{ij} + |\Omega_i| S_{O_i}(U) + |\Omega_i| S_{f_i}(U). \tag{11}$$

We then use the rotational invariance between $G$ and $H$ around each side (Toro (2001); Loukili and Soulaimani (2007); Hemker and Spekreijse (1985)), which gives

$$H(U).n_{ij} = T_{n_{ij}}^{-1} G(T_{n_{ij}} U), \quad T_{n_{ij}} = \begin{bmatrix} 1 & 0 & 0 \\ 0 & n_{ij}^1 & n_{ij}^2 \\ 0 & -n_{ij}^2 & n_{ij}^1 \end{bmatrix}. \tag{12}$$

In this approach we use a finite volume method centered on the cells with piecewise constant variables. To approximate the fluxes, we solve a unidirectional Riemann problem in the direction $n_{ij}$, which allows (11) to be written as

$$|\Omega_i| \frac{dU_i}{dt} = -\sum_{j=1}^{3} L_{ij} T_{n_{ij}}^{-1} \tilde{G}(T_{n_{ij}} U_i, T_{n_{ij}} U_j) + |\Omega_i| S_{O_i}(U) + |\Omega_i| S_{f_i}(U), \tag{13}$$

with $\tilde{G}(T_{n_{ij}} U_i, T_{n_{ij}} U_j) = \tilde{G}(U_L, U_R)$ a discrete flux found by solving a Riemann problem with $U_L = T_{n_{ij}} U_i$ and $U_R = T_{n_{ij}} U_j$ as the initial states. That is to say,

$$\begin{cases} \dfrac{\partial U}{\partial t} + \dfrac{\partial G(U)}{\partial x_n} = 0, \\[2mm] U(x,0) = \begin{cases} U_L \text{ if } x_n < 0, \\ U_R \text{ if } x_n > 0. \end{cases} \end{cases} \tag{14}$$

As far as the boundary conditions are concerned, we choose to proceed as in Loukili and Soulaimani (2007). A transmissive condition is solved by supposing a state $U_R = U_L$ in the resolution of the Riemann problem. For a condition with an incoming flow $Q$ we calculate the flow directly with $\tilde{G}(U_L) = (Q, Q^2/h_l + (gh_l)^2/2, 0)^T$, and for a nontransmissive wall condition we use the preceding calculation with a zero incoming flow, ie $\tilde{G}(U_L) = (0, (gh_l)^2/2, 0)^T$.

Temporal discretization is done using an explicit Euler method, which makes it possible to avoid having to solve a linear system at the cost of a time step constrained by a stability condition. The stability analysis from Loukili and Soulaimani (2007) gives the following CFL condition

$$CFL = \Delta t \frac{max(\sqrt{gh} + \sqrt{u^2 + v^2})}{min(d_{L,LR})}, \tag{15}$$

with $d_{L,LR}$ the distance between the cell center and the L/R interface. However, for simplicity we choose to take $d_{L,LR} = R_L$, the radius of the circle inscribed in cell L. This is a very conservative condition and has proven to result in great stability for CFL = 0.9.

Using the explicit Euler discretization, (13) becomes

$$\frac{U_i^{n+1} - U_i^n}{\Delta t} = -\frac{1}{|\Omega_i|} \sum_{j=1}^{3} L_{ij} T_{n_{ij}}^{-1} \tilde{G}(U_L^n, U_R^n) + |\Omega_i| S_{O_i}^n(U) + |\Omega_i| S_{f_i}^n(U), \tag{16}$$

where the $n$ exponent shows that the values are taken at times $t_n$.

## 3.2. HLL and HLLC schemes

We recall here the flux of the HLL scheme (Loukili and Soulaïmani, 2007; Toro, 2001; Harten et al., 1983)

$$\tilde{G}^{HLL} = \begin{cases} G(U_L), & S_L \geq 0, \\ G(U_R), & S_R \leq 0, \\ \tilde{G}(U_*), & S_L \leq 0 \text{ and } S_R \geq 0, \end{cases} \tag{17}$$

with $\tilde{G}(U_*)$ the flux in the star region given by

$$\tilde{G}(U_*) = (S_R G(U_L) - S_L G(U_R) + S_R S_L (U_R - U_L))/(S_R - S_L). \tag{18}$$

The right and left wave speeds, $S_R$ and $S_L$, are estimated as follows

$$S_L = u_L - a_L p_L, S_R = u_R + a_R p_R, \tag{19}$$

where $k = L, R$, $a_k = \sqrt{gh_k}$ and

$$p_k = \begin{cases} [h^*(h^* + h_k)/2)]^{1/2}/2, & h^* > h_k, \\ 1, & h^* \leq h_k, \end{cases} \tag{20}$$

with $h^*$ the water height in the star region.

The water height $h^*$ is evaluated in multiple steps. A first approximation indicates if this is a shock wave or a rarefaction wave

$$h_0^* = \frac{h_L + h_R}{2} - \frac{(U_R - U_L)(h_L + h_R)}{4(a_R + a_L)}. \tag{21}$$

If $h_0^* \leq min(h_L, h_R)$, then it is a rarefaction wave and we have

$$h^* = [(a_R + a_L)/2 + (U_L - U_R)/4]^2/g, \tag{22}$$

whereas, if $h_0^* > min(h_L, h_R)$, then it is a shock wave and we have

$$h^* = (h_L g_L + h_R g_R + u_L - u_R)/(g_L + g_R),$$
$$g_k = \left[ \frac{g(h_0^* + h_k)}{2h_0^* h_k} \right]^{1/2}, k = L, R. \tag{23}$$

We can then modify this scheme to obtain HLLC (Loukili and Soulaïmani, 2007; Ata et al., 2013; Toro, 2001; Harten, 1983) by accounting for the speed $S^*$ in the star region. This step only modifies the last component of the flux, as follows

$$\tilde{G}_3^{HLLC} = \begin{cases} \tilde{G}_1^{HLL}(U_L, U_R)v_L, & S^* \leq 0, \\ \tilde{G}_1^{HLL}(U_L, U_R)v_R, & S^* \geq 0, \end{cases} \tag{24}$$

with

$$S^* = \frac{s_L h_R(u_R - S_R) - s_R h_L(u_L - S_L)}{h_R(u_R - S_R) - h_L(u_L - S_L)}, \tag{25}$$

by considering the following dry bed situations

$$\begin{aligned} h_L = 0 &: S_L = U_R - 2a_R, & S_R = u_R + a_R, S^* = S_L, \\ h_L = 0 &: S_L = U_L - a_L, & S_R = u_L + 2a_L, S^* = S_R. \end{aligned} \tag{26}$$

With this new flux, (16) becomes

$$\frac{U_i^{n+1} - U_i^n}{\Delta t} = -\frac{1}{|\Omega_i|} \sum_{j=1}^{3} L_{ij} T_{n_{ij}}^{-1} \tilde{G}^{HLLC}(U_L^n, U_R^n) + S_{O_i}^n(U) + S_{f_i}^n(U) \tag{27}$$

### 3.2.1. Bathymetric source term

To compute the bathymetric source term, we use the method in Ata et al. (2013) that is based on Audusse and Bristeau (2005). It is a method that can be used with any consistent numerical scheme (Ata et al. (2013)).

The main steps are:

- Define the interface bathymetry between cells i and j with $z_{ij} = z_{ji} = max(z_i, z_j)$ .

- Define the water depth at the interface $h_{ij}^{**} = max(0, h_i + z_i - z_{ij})$ and redefine the interface unknowns

$$U_{ij}^{**} = (h_{ij}^{**}, h_{ij}^{**} u_i, h_{ij}^{**} v_i)^T. \tag{28}$$

- Use the hypothesis $\nabla s \simeq 0$; so that $-gh\nabla b \simeq \nabla(gh^2/2)$, which leads to

$$S_O(U) = \begin{pmatrix} 0 \\ \nabla(gh^2/2) \end{pmatrix}. \tag{29}$$

Then, by using the definition in (10), we choose a new discretization with the new set of unknowns in (28) to get (Audusse et al., 2004b)

$$S_{O_i}(U) = \frac{1}{|\Omega_i|} \sum_{j=1}^{3} L_{ij} \begin{pmatrix} 0 \\ g(h_{ij}^{**2} - h_i^2)n_{ij} \end{pmatrix}. \tag{30}$$

- Finally, we replace $S_{O_i}$ in (27) with its expression in (30).

### 3.2.2. Friction

We recall that

$$S_{f_i}(U) = \frac{1}{|\Omega_i|} \int_{\Omega_i} S_f(U) \, \partial\Omega, \tag{31}$$

with

$$S_f(U) = \left(0, -gh\frac{m^2\bar{u}\sqrt{\bar{u}^2 + \bar{v}^2}}{h^{4/3}}, -gh\frac{m^2\bar{v}\sqrt{\bar{u}^2 + \bar{v}^2}}{h^{4/3}}\right). \tag{32}$$

To deal with this nonlinear term, we take up the semi-implicitation proposed in Loukili and Soulaïmani (2007). This method consists of taking

$$S_f = \frac{S_f^{n+1} + S_f^n}{2}, \tag{33}$$

and then making the approximation

$$S_f^{n+1} \simeq S_f^n + J_f(U^{n+1} - U^n), \tag{34}$$

with

$$J_f^n = \frac{\partial S_f^n}{\partial U} = \begin{pmatrix} 0 & 0 & 0 \\ \partial S_{fx}^n / \partial h & \partial S_{fx}^n / \partial(h\bar{u}) & \partial S_{fx}^n / \partial(h\bar{v}) \\ \partial S_{fy}^n / \partial h & \partial S_{fy}^n / \partial(h\bar{u}) & \partial S_{fy}^n / \partial(h\bar{v}) \end{pmatrix}. \tag{35}$$

Equation (31) can then be approximated by

$$S_{f_i}(U) = S_f(U_i). \tag{36}$$

Finally, replacing (33) in (27) gives the final form of the discretization

$$\frac{U_i^{n+1} - U_i^n}{\Delta t} = \left[ I - \frac{\Delta t}{2} J_f \right]^{-1} \left[ -\frac{1}{|\Omega_i|} \sum_{j=1}^{3} L_{ij} T_{n_{ij}}^{-1} \tilde{G}^{HLLC}(U_L^n, U_R^n).n_i + S_f(U_i^n) + S_{O_i}^n(U_i^n, U_{ij}^{**n}, n_{ij}) \right]. \tag{37}$$

## 4. Domain Decomposition

In this section, we present the pre-processing steps implemented in order to perform the domain decomposition. The objective is to generate a mesh file for each sub-domain, starting from the mesh file of the whole domain. We use a fairly classical method, diagrammed in Figure 2. All of the major domain decomposition steps are presented in the following subsections.

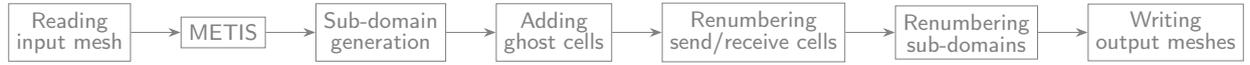

**Figure 2:** Pre-processing of the mesh files using METIS

### 4.1. Using METIS

To perform the domain decomposition, we use the METIS library (Karypis and Kumar, 2009). METIS is a widely used library that allows a mesh to be decomposed into sub-domains using the mesh's graph's partitioning. This step is crucial, as it is necessary to limit the contact surfaces between the sub-domains as much as possible in order to have the minimum amount of memory exchanged between the processors. METIS does this very quickly and efficiently.

We use METIS to attribute a sub-domain to each cell. Next, we add one layer of ghost cells on each sub-domain. This process is described in the following section.

### 4.2. Adding ghost cells

Following the decomposition of the initial mesh, we obtain the number of desired sub-domains. We manage these sub-areas by anticipating what will be needed during the memory exchanges.

As presented in section 2, our numerical scheme requires knowing the unknowns in the neighboring cells, i.e., the cells that have an edge common to the computed cell. This requirement implies the need for special treatment at the edges of the sub-domains.

The cells on the edges of the original domain are treated with the boundary conditions presented in section 2. Special care must be given to the cells on the edges of the sub-domains that are in contact with other sub-domains. To calculate the new value of a cell on such an edge, we determine the value of the neighboring cells that are potentially in another sub-domain and thus allocated to another processor. This is when we will have to use the MPI library to perform a memory exchange between the processors.

To avoid having to perform a memory exchange each time we try to compute a border cell, a common approach is to add a layer of so-called ghost cells to each sub-domain. Thus, the values of all these cells can be recovered at one time, with no further issues about memory exchanges during the next time step's computation.

To find the ghost cells, the first step is to find the nodes that are common between two adjacent sub-domains. Figure 3(a) shows the allocation of the cells made by METIS; one area is colored in blue, the other in white. We show the dividing line between the domains in red. The nodes common to the two sub-domains are on this red line.

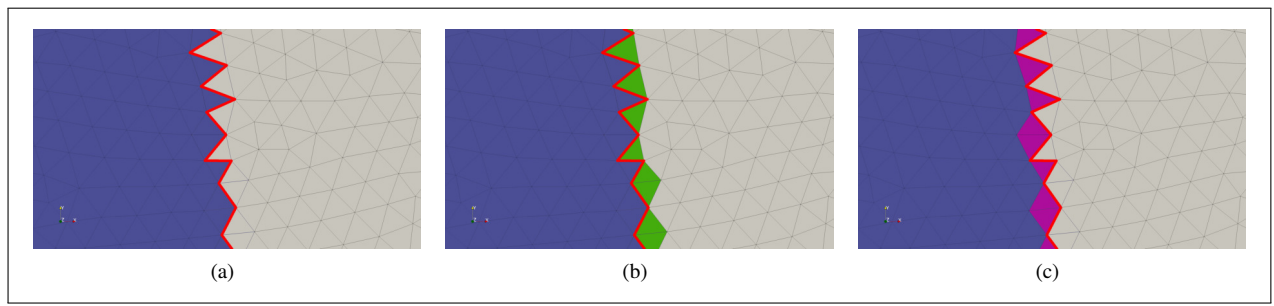

**Figure 3:** Allocation of cells made by METIS at the interface between two sub-domains. The blue cells correspond to one domain, the white cells to the other domain. The line of nodes common to the two sub-domains is represented in red. Green cells are added to the blue sub-domain, and purple cells are added to the white sub-domain.

After finding these common nodes, each sub-domain is examined to determine the cells with two successive nodes that belong to the list of the common nodes. These will be ghost cells for the other sub-domain. Figure 3(b) shows the ghost cells of the blue sub-domain in green. They will be received by the blue sub-domain and sent from the white sub-domain. Similarly, Figure 3(c) shows the ghost cells of the white domain in purple. They will be received by the white sub-domain and sent from the blue sub-domain.

As it is only done once for each mesh, this serial domain decomposition is good enough for our purposes. In the future, this decomposition may need to be upgraded by using parallel computing, as in Patchett et al. (2017).

### 4.3. Renumbering the ghost cells and generating the sending/receiving information

The numbering of the cells that will be sent/received is essential for the memory exchange's performance. Indeed, when a message is sent with MPI, there is a latency time. Hence, messages should be grouped as much as possible to send a single large one rather than several small ones.

Our goal is to send all the cells needed by an adjacent sub-domain at one time by sending a memory block. Thus, the cells need to be grouped in a single block by our numbering. We will therefore renumber the cells to have: first, the cells that will neither be sent nor received; then the cells to be sent to other sub-domains, grouped in as many blocks as there are adjacent sub-domains; and finally the ghost cells that must be received, gathered in as many blocks as there are adjacent sub-domains.

Considering how the ghost cells are added to each sub-domain, it is natural that these cells are located at the end of the numbering and form a block for each adjacent sub-domain. On the other hand, the cells that each sub-domain must send are derived from the mesh's initial numbering and are therefore not generally side-by-side in the numbering. Therefore, we will re-number these cells to ensure that they are correctly grouped into blocks. The objective is to have a mesh file as shown in Figure 4.

Figure 4: Mesh file for the 4th sub-domain.

This process allows us to generate directly in the mesh files all the information that will be necessary to send and receive messages by block in the simulation. In particular, as presented in Figure 4, each mesh file will contain the starting indices of each block to be sent, their size, and the index of the sub-domain to which the block must be sent. The process works in the same way for the cells to receive; each mesh file contains the starting index of the block of ghost cells to receive, the size of this block, and the number of the sub-domain from which this block will be received.

In figure 4, each cell category is associated with a color:

- Uncolored cells are neither to send nor to receive;
- Red cells are to be sent to sub-domain 3;
- Orange cells are to be sent to sub-domain 5;
- Cyan cells are to be received from sub-domain 3;
- Purple cells are to be received from sub-domain 3.

Generating this data in pre-processing thus simplifies the task in the simulation code, as all the information necessary to correctly carry out the ghost cells' exchange can be read in the meshes' files.

## 4.4. Sub-domain renumbering
The goal here is to renumber the sub-domains of the decomposition, meaning the renumbering of the sub-domains themselves rather than a renumbering of the cells in each sub-domain.

For most computer clusters, there are usually 4 to 8 GPUs per computer node. If we plan to use 4 GPUs per node, we will have four sub-domains per node, and therefore we would have to optimize the renumbering accordingly. However, there is often no quick fix that would avoid all of the exchanges between nodes.

This renumbering does not have a significant impact on the code's performance; in the best case, it improved the performance by $8 - 10\%$. We present it here because it is very straightforward, simple to implement and could be especially useful when dealing with many sub-domains.

While METIS already has a way to deal with the numbering of sub-domains, we found that in our case, it could easily be optimized. To improve this numbering, we propose to use the well-known Cuthill-McKee algorithm (Cuthill and McKee, 1969).

In the classic Cuthill-McKee algorithm, we choose to start numbering with an edge domain by looking for the domain with the least important degree, i.e., the domain with the fewest neighbors. However, in our case, where there are only a few sub-domains, many have only two neighbors, which complicates being able to start the numbering with an edge domain. To solve this problem as simply as possible, we suggest starting the numbering with a domain that contains input nodes from the original domain. These nodes are given in our meshes, and are then used to define the boundary conditions. Therefore, it becomes a simple task to identify the sub-domains that contain input nodes.

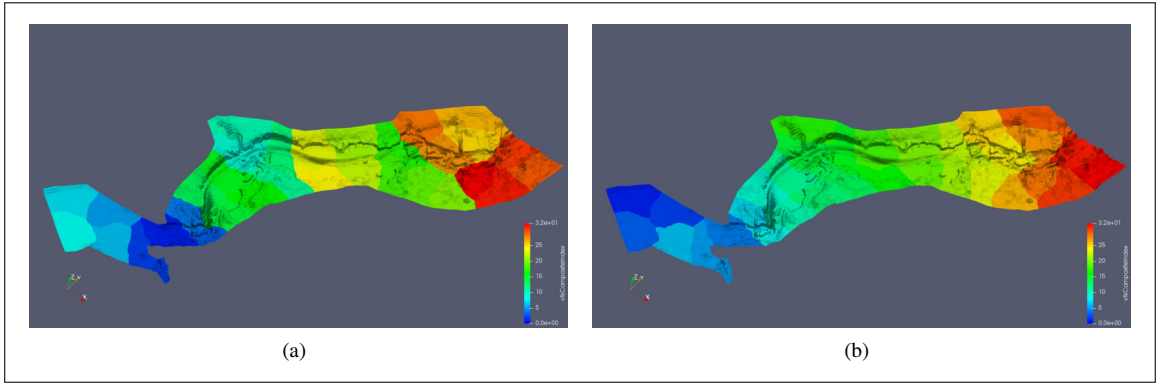

(a)                                          (b)

**Figure 5:** Domain of the Mille Îles river broken into 32 sub-domains, each corresponding to a color. The METIS numbering is on the left (a), our renumbering is on the right (b).

Once this first domain has been selected, we can use the classic Cuthill-McKee algorithm. As an example, in Figure 5 we show a comparison of the numbering of 32 sub-domains on the Mille Îles river.

As expected with the Cuthill-McKee algorithm, the spatially-close sub-domains have close numbers, which implies that they will probably be on the same computation node. However, as we have specified, there will be exchanges between domains on separate nodes no matter how the renumbering is done.

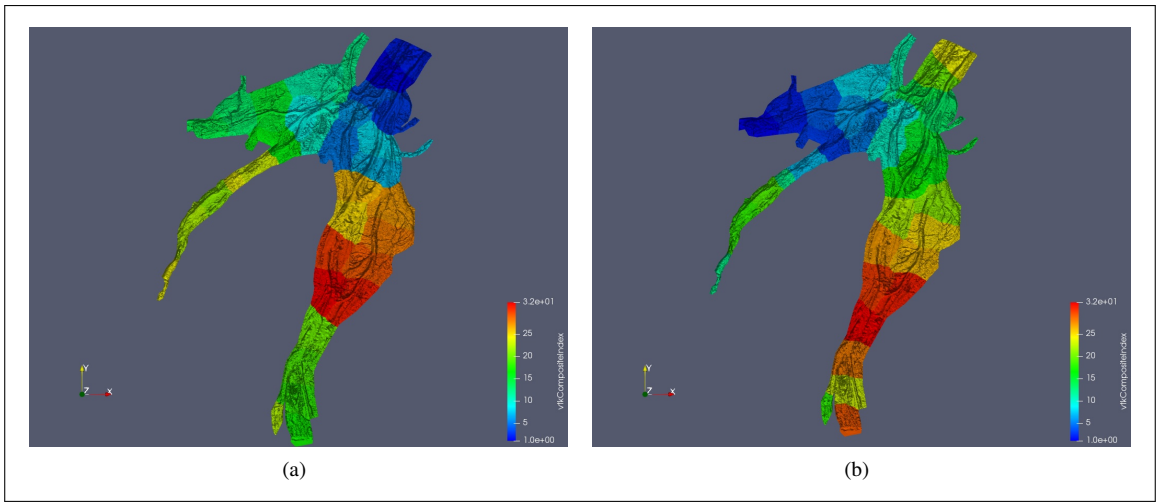

(a)                                          (b)

**Figure 6:** Domain of the Montreal archipelago broken into 32 sub-domains, each sub-domain corresponds to a color. The METIS numbering is on the left (a), our renumbering is on the right (b).

Although there is a rather satisfactory renumbering on the Mille Îles river domain presented in Figure 5, this method quickly reaches its limits with domains containing several entries, such as the domain of the Montreal archipelago. The METIS numbering shows these limits on the left in Figure 6. As with the renumbering on the Mille Iles river, Figure 6 shows close sub-domains that have distant numbers.

Our renumbering is on the right of Figure 5. In this case, it is difficult to say whether our numbering is better than the METIS numbering on the left. This is confirmed by the small performance improvement of $2 - 3\%$.

If the objective is to treat more and more domains containing several inputs and outputs, and to break them down into even more sub-domains, then a better method for this renumbering will have to be proposed. For our purposes, this simple renumbering is sufficient.

As stated earlier, this renumbering of the sub-domains led to, at most, a 10% speed-up of the computation on the Mille Îles river decomposed into 32 sub-domains.

## 5. CUDA, MPI and CUDA-Aware MPI

GPU programming in CUDA is discussed here to give the reader some background to understand how it works with MPI. For further information about GPU programming see Sanders and Kandrot (2010), Fatica and Ruetsch (2014) and Kirk and mei W. Hwu (2017).

The general idea of CUDA is to use the CPU to launch specially-written functions on the GPU. Such functions are called *kernels* and are launched in parallel on the GPU following a specified configuration of threads and blocks. It must be noted that the *kernels* only work with variables that are on the GPU. In CUDA-Fortran, a variable should be declared with the *device* attribute in order to be allocated on the GPU. As the variables are declared on the GPU, a copy from the host memory (CPU memory) to the device memory (GPU memory) will most often be made just before the calling of a kernel.

Thus, the goal is to reduce the number of exchanges between the CPU and the GPU in order to speed up the computations. In our case, we wrote all the functions used in the time loop of the algorithm on the GPU. This allows us to initialize and set up all the vectors on the CPU side at the beginning, then copy all of these vectors on the GPU, perform the iterations over time on the GPU without any exchanges between the CPU and the GPU, and finally copy back the vectors from the GPU to the CPU to do the post-processing.

MPI (Message Passing Interface) is a standard that defines the communication functions between several processors or remote computers. We used OpenMPI (Gabriel et al., 2004) in order to run different copies of the code on different processors. Each MPI process thus has its own memory, which will be the only one it is able to modify. Furthermore, each MPI process will be associated with a unique GPU on which it will be the only one to launch *kernels* on.

### 5.1. Classic GPU memory exchange

Usually, MPI only allows memory exchanges between variables defined on the CPU. When we couple MPI with CUDA-Fortran, each MPI process has its own variables on the CPU and on the GPU it is associated with. To exchange the variables of the GPU, it is necessary, in the classic case, to copy these variables onto the CPU, then make the MPI memory exchange, and finally copy the variables on the GPU in return. These copies are done using the *cudaMemcpy* function from CUDA.

For example, to make the classic reduction on the time step *dt*, one can do the following,

```
ierr = cudaMemcpy(dt,dt_d,1)
call MPI_ALLREDUCE(MPI_IN_PLACE, dt, 1,
       fp_kind_mpi, MPI_MIN, MPI_COMM_WORLD, mpi_ierr)
ierr = cudaMemcpy(dt_d,dt,1)
```

where $dt\_d$ has been declared on the GPU using the attribute *device*.

There are more complex ways to overlap memory exchanges with calculations, such as by using CUDA *streams* to make an asynchronous copy between the CPU and the GPU and by using the non-blocking version of *ALLREDUCE*, *IALLREDUCE*.

Although the latter method works, it is generally not the most effective, especially if the exchanges are not overlapped with calculations. A better method, which is also simpler to program, is to use CUDA-Aware OpenMPI, as detailed below.

### 5.2. GPU Memory exchange using CUDA-Aware OpenMPI

A CUDA-Aware OpenMPI version indicates that the OpenMPI library has been compiled with support for CUDA. Such a compiled library makes it possible to send and receive GPU memory directly without copying the GPU memory onto the CPU. This possibility has been available since version 1.7 of OpenMPI.

In the CUDA-Aware OpenMPI documentation, it states, "Now, the Open MPI library will automatically detect that the pointer being passed in is a CUDA device memory pointer and do the right thing" (`https://www.open-mpi.org/faq/?category=runcuda`). We are not going to detail what *the right thing* means but we can give offer some clues.

If two GPUs are on the same computer node, they may communicate in Peer-To-Peer, i.e., they can communicate directly with each other without going through the host's memory. We can then write a function that checks whether the GPUs can communicate in Peer-to-Peer. If this is the case, we can perform a memory exchange directly between the two GPUs using the *cudaMemcpy2D ()* function of CUDA Fortran. If the GPUs cannot communicate in Peer-to-Peer, we will be forced to copy the GPU variables to the CPU and then use MPI to exchange memory. This is the maximum that we can do in terms of programming; using CUDA-Aware OpenMPI allows us to avoid this situation and instead use the exchange in Peer-to-Peer whenever possible.

It is clear that CUDA-Aware OpenMPI lightens the programming task load, allowing us not to be concerned about how the memory exchange is done. However, as certain functionalities are too low-level to be accessible from CUDA-Fortran, the use of CUDA-Aware OpenMPI becomes necessary to achieve the best performances. One such functionalities is the GPU RDMA (Remote Direct Memory Access) which allows direct memory exchange between GPUs on different computation nodes. This functionality is not accessible to programming in MPI and CUDA-Fortran, the only way to use it is through a CUDA-Aware version of OpenMPI. These are obviously not the only two advantages of using CUDA-Aware OpenMPI; as the library manages everything, all unnecessary memory exchanges are avoided.

Using CUDA-Aware OpenMPI also has some disadvantages. For example, it is more difficult to know what is used by the library to perform a memory exchange, and it can be difficult to determine if the system is really being used at its maximum level. In addition, for proper operation, the OpenMPI library must be particularly compiled to accommodate the system. In our case, it must be compiled with support for CUDA and for the *pgf90* compiler.

Using CUDA-Aware OpenMPI, the reduction on the time step shown earlier can be directly performed on the variable *dt_d* in the following way:

```
call MPI_ALLREDUCE(MPI_IN_PLACE, dt_d, 1,
        fp_kind_mpi, MPI_MIN, MPI_COMM_WORLD, mpi_ierr)
```

The same approach works for the functions *SEND* and *RECV*, which are used to communicate the ghost cells at each step. This step is easy to perform, as the file format presented in Figure 4 gives us all the information we need to send and receive the ghost cells as blocks of data.

### 5.3. Overlapping MPI memory exchanges with computation in our in-house code CuteFlow

Here we present how we overlapped the computations with the MPI memory exchanges in **CuteFlow**, our in-house solver for the SWE.

**CuteFlow** is an in-house research code for solving the SWE with the finite volume scheme described in section 2. The first sequential version on a CPU was developed by Azzeddine Soulaïmani and Youssef Loukili (Loukili and Soulaimani, 2007), and subsequently adapted by Jean-Marie Zokagoa (Zokagoa and Soulaïmani, 2010). Arun Kumar Suthar Suthar and Soulaimani (2018) then used CUDA Fortran to make use of a single GPU. Figure 7 shows the inner processes of the **CuteFlow** solver which are very common among time-explicit finite volume solvers.

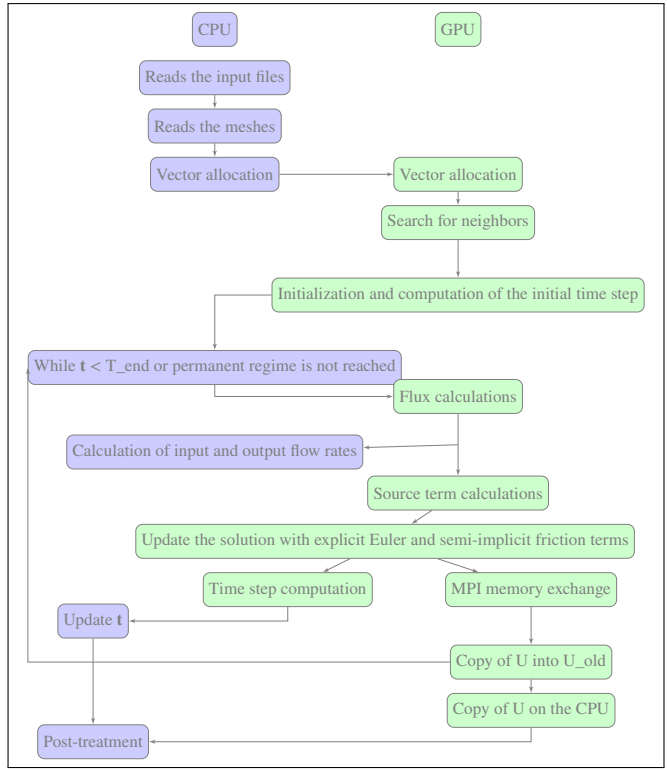

Figure 7: Flow chart of **CuteFlow**

### 5.3.1. Compilation
**CuteFlow** uses MPI and CUDA-Fortran. To compile, we use the OpenMPI wrapper *mpif90* with the PGI compiler *pgf90*. The following versions of different modules are used on the computer clusters:

- pgi/19.4 ;
- cuda/10.0.130 ; and
- openmpi/3.1.2 .

More information on a CUDA-Aware version of Open-MPI can be found on the official website `https://www.open-mpi.org/faq/?category=runcuda`.

### 5.3.2. Overlapping
### computations and memory exchanges
When using MPI and CUDA-Fortran there are many ways to overlap computations and memory exchanges. One approach is to use CUDA *streams* to overlap CPU-GPU memory exchanges with computations on the GPU. For example, when calculating the inflow and outflow rates, data can be sent from the GPU to the CPU on a particular CUDA *stream* so that the GPU computations can continue without interruption.

In order to get the best performance, MPI memory exchanges must be overlapped with computations. In our code, it is simple to compute the CFL time step while performing MPI memory exchanges asynchronously with the *ISEND* and *IRECV* functions. Thanks to the use of a CUDA-Aware version of OpenMPI, these functions allows us to exchange GPU memory asynchronously by using CUDA *streams* internally.

As soon as the new solution has been updated, the new time step's computation can begin. This computation is not likely to take very long on its own, as each thread will calculate a local time step for each cell. However, when we add the reduction performed on the GPU (see Harris (2007)) to find the minimum time step in the sub-domain, and then the reduction via an *ALLGATHER* on the time step to find the minimum time step of all the sub-domains, the computation becomes quite long. We therefore take this opportunity to exchange ghost cells between the sub-domains at the same time. This exchange does not pose a problem because the computation of the new time step only depends upon the sub-domain's interior cells. We launch the non-blocking MPI exchanges of ghost cells just before performing the time step computations so that these two steps overlap as much as possible.

The impact of overlapping computations with the MPI memory exchange is discussed in section 6, as it leads to far better results than using blocking exchanges.

# 6. Results

After presenting the computer cluster we utilized, and defining how we calculate speed-up and efficiency, here we present some results for the case of a dam-break flow in a flat canal, for the Mille Îles river and finally for the Montreal archipelago.

## 6.1. Computer cluster used

The results presented in this section were produced using BELUGA, a cluster managed by *Compute Canada* and *Calcul Quebec*. At the time of this writing it contained 172 GPU nodes consisting of 2 Intel Gold 6148 Skylake @ 2.4 GHz and 4 NVidia V100SXM2 (16G memory), connected via NVLink. The full configuration can be found on the *Compute Canada* website `https://docs.computecanada.ca/wiki/B%C3%A9luga/en`.

## 6.2. Definitions of speed-up and efficiency

We recall the definitions of speed-up and efficiency. Most of the time, we compare the times on $n$ GPUs with the times on one GPU. Thus, for the speed-up, we have

$$\text{Speed-Up} = \frac{\text{Time on 1 GPU}}{\text{Time on } n \text{ GPU}},$$

and for the efficiency

$$\text{Efficiency} = \frac{\text{Time on 1 GPU}}{n * \text{Time on } n \text{ GPU}}.$$

## 6.3. Case of one-dimensional dam failure

The test case presented here corresponds to the resolution of a one-dimensional Riemann problem. To simulate this problem, we take a rectangular domain $\Omega = [-10, 10] * [0, 100]$ and choose the following initialization values

$$\begin{cases} h & = \begin{cases} 10 \text{ if } y < 50, \\ 1 \text{ if } y > 50. \end{cases} \\ \bar{u} & = 0 \\ \bar{v} & = 0. \end{cases}$$

We use a basic mesh that we generate from points placed on a grid in the xy plane. Figure 8 shows a coarse mesh of a domain with 400 elements. The results presented in the following sub-sections are computed on much finer meshes, ranging from 400,000 to 13,000,000 elements.

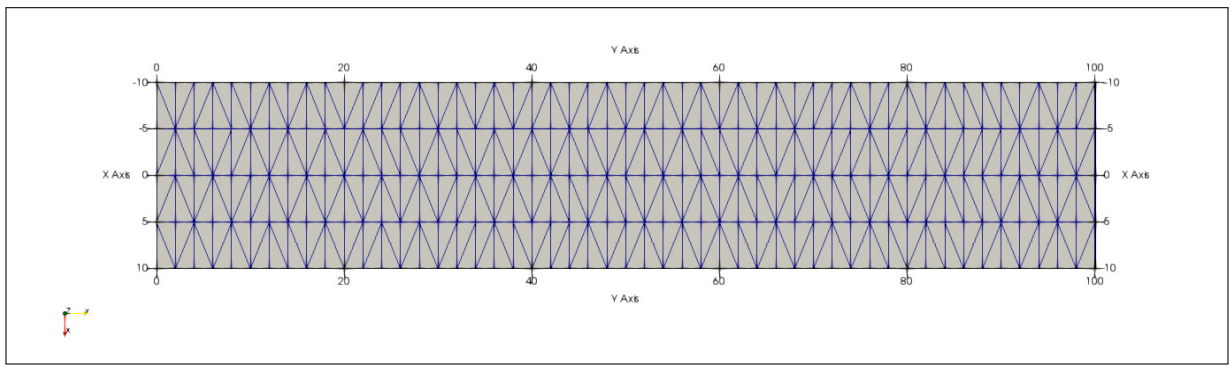

**Figure 8:** 400-cell mesh for the one-dimensional dam-break case

### 6.3.1. Solutions

The solutions presented in Figures 6.3.1 and 6.3.1 were computed using our in-house code on a 400,000-cell mesh. These are very classic results that can be found in Toro (2001); Ata et al. (2013); Zokagoa and Soulaïmani (2010).

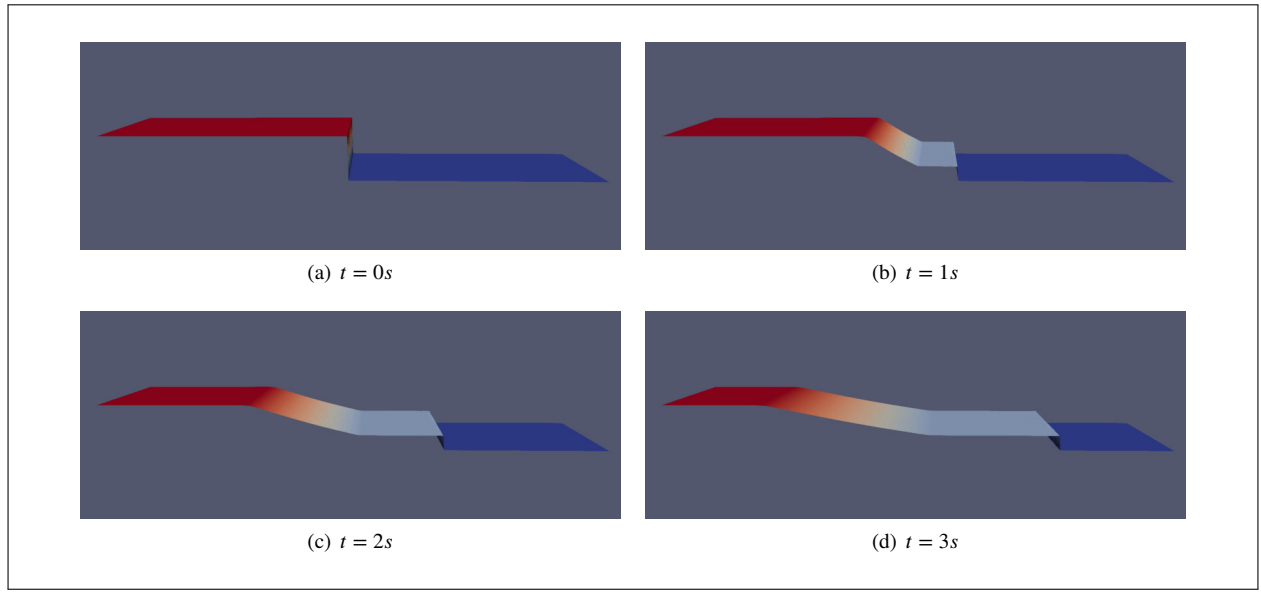

(a) $t = 0s$　　　　(b) $t = 1s$

(c) $t = 2s$　　　　(d) $t = 3s$

**Figure 9:** Solutions for the one-dimensional dam-break case on a 400,000-cell mesh at different times.

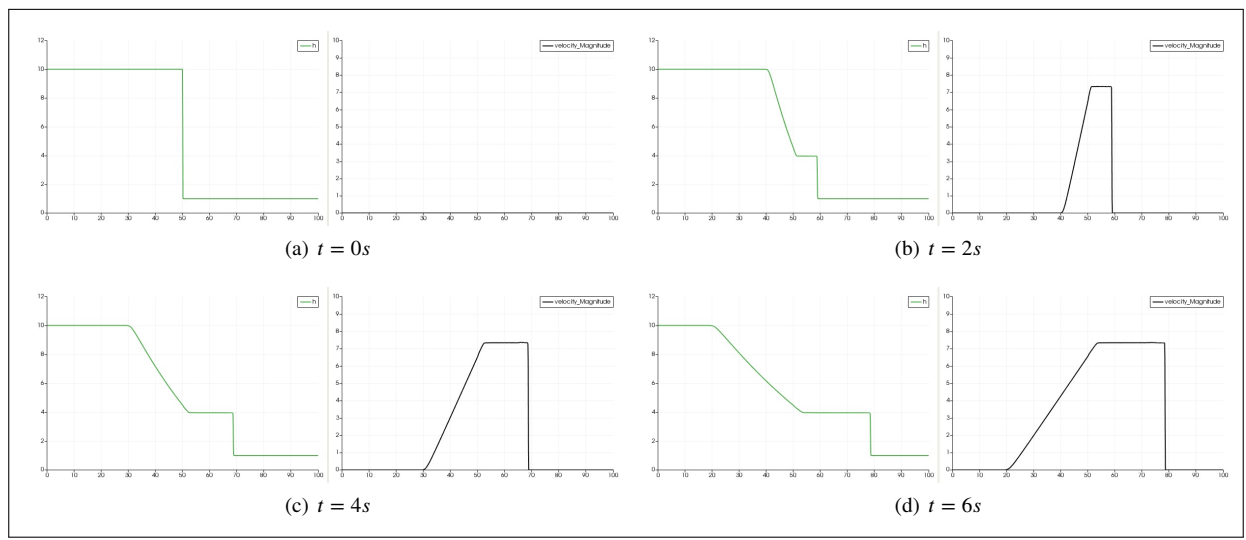

(a) $t = 0s$　　　　(b) $t = 2s$

(c) $t = 4s$　　　　(d) $t = 6s$

**Figure 10:** Projected solutions on the y-axis for the one-dimensional dam-break case on a 400,000-cell mesh. For each time, water depth is plotted on the left and velocity magnitude is plotted on the right.

### *6.3.2. Speed-up and efficiency for different meshes*

Table 1 presents the speed-up and efficiencies obtained for various meshes of 400,000, 1,600,000, 6,300,000 and 13,000,000 elements.

First, we can see that the performance of the non-blocking exchange is always much better than that of the blocking exchange. This is as we expected, because the non-blocking exchange allows us to launch all memory exchanges at the same time in addition to superimposing them on the calculations, such as the computation of the CFL condition.

Next, we note that on the smallest mesh of 400,000 elements, we are far from the ideal speed-up with a maximum acceleration factor of 4, even using 16 GPUs. This can be explained quite simply: the mesh is too small to allow the calculations to be superimposed correctly on the memory exchanges. In this case, we are strictly limited by latencies, whether they come from the launch of kernels on the GPUs or from MPI memory exchanges. We can also observe that the larger the mesh becomes, the closer the speed-up is to the ideal.

From these results, we can determine the optimal number of elements per GPU. We choose here to consider optimal as an efficiency greater than 80%. With this choice, the optimal number of elements per GPU appears to be between 300,000 and 500,000 elements. This means that to obtain the results that we consider optimal, we must use 1 GPU for the case of 400,000 elements, 4 GPUs for the case of 1,600,000 elements, 12 GPUs for the case of 6,300,000 elements, and 20 GPUs for the case of 13,000,000 elements.

Obviously, this choice of an efficiency greater than 80% is arbitrary. During the first phase of the simulations, when we are trying to have the first stabilized solution, we may want to proceed as quickly as possible without worrying about efficiency. It is not a problem to have a low efficiency during the first phase, because we only launch a single simulation, and thus only use a few resources on the computation clusters. On the other hand, during the second phase, where we build the simulation database, we will need an efficiency greater than 80%, as we will be launching hundreds of simulations. This will be the most costly phase of resource use on the calculation clusters.

## 6.4. Case of the Mille Îles River

Here we present results obtained on the Mille Îles river domain. On Figure 11 we can see the simulation domain superimposed on a satellite image recovered by Google Earth Pro. The image is oriented with the y-axis to the north. The entrance to the domain is at the bottom left of the image and the exit is at the top right.

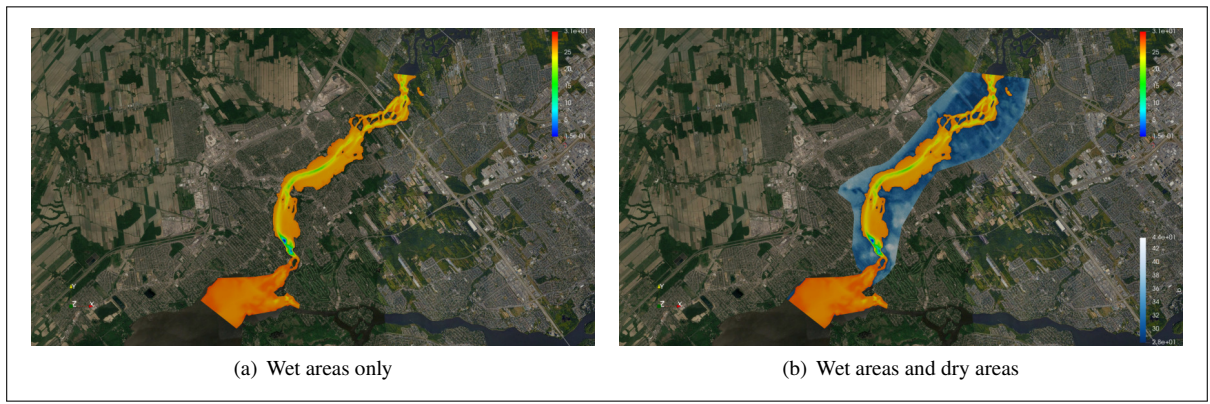

(a) Wet areas only                    (b) Wet areas and dry areas

**Figure 11:** Mille Îles river domain overlapped with a satellite image from Google Earth Pro.

We have several versions of meshes for this domain which vary between 200,000 and 11 million elements. The mesh is refined in the critical zones, in particular around the piers of a bridge that crosses the river. Figure 12 shows the area of the dam under the first bridge of the domain with a high degree of refinement.

**Table 1**
Speed-up and efficiencies for different types of meshes for the dam-break test case

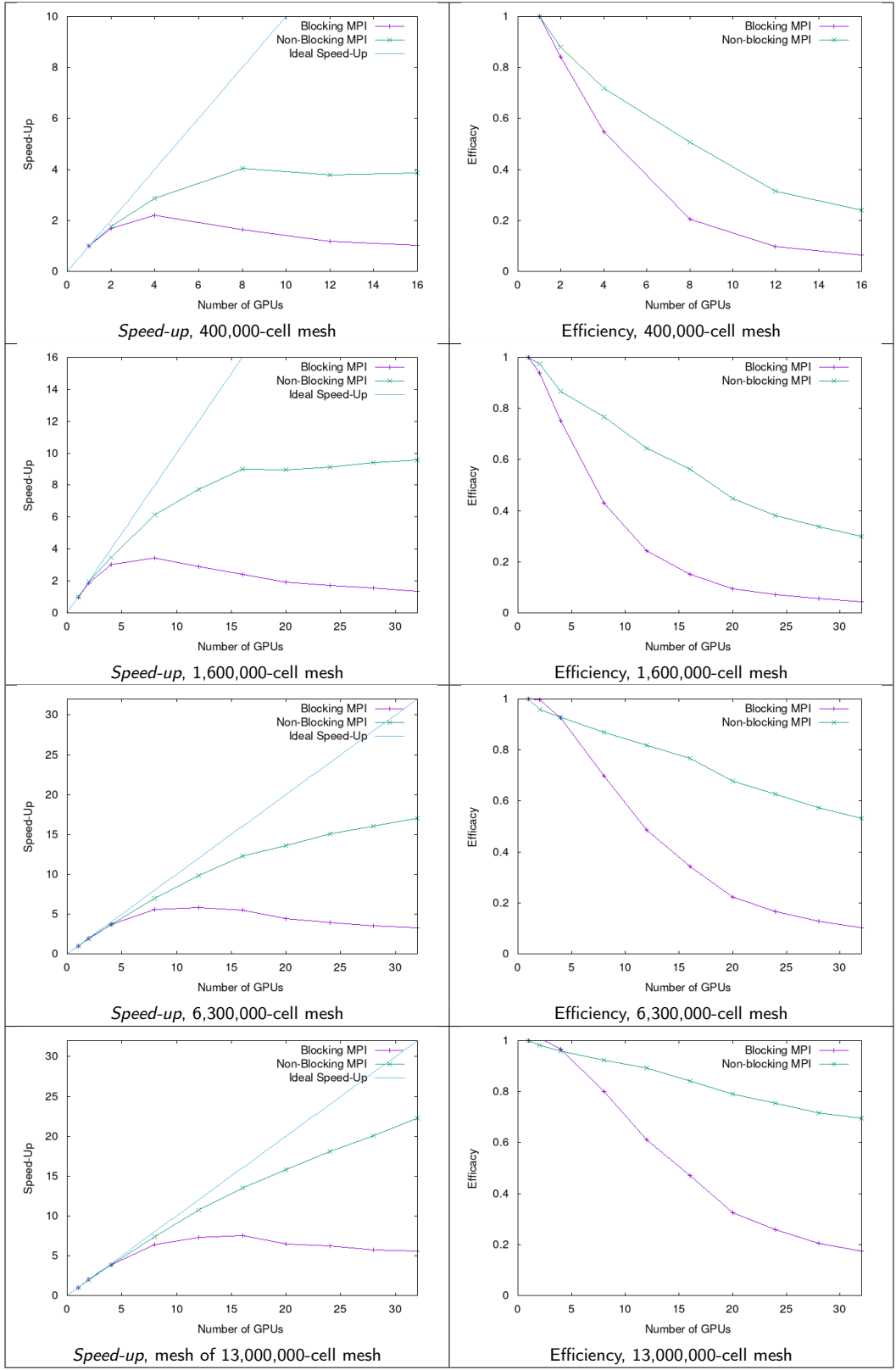

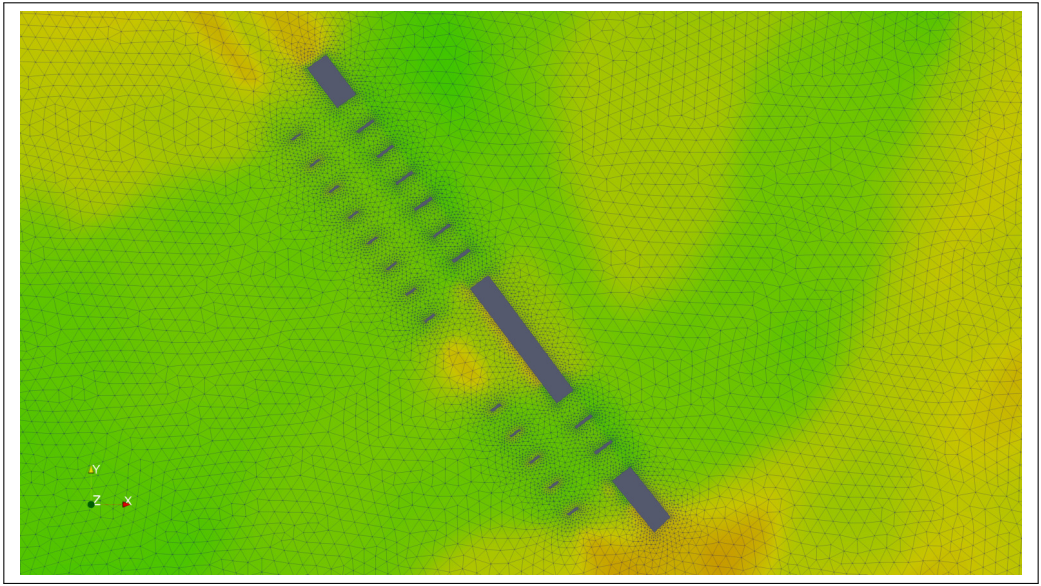

**Figure 12:** Refinement zone around the piers of a bridge in the Mille Îles river, 200,000-cell mesh.

### 6.4.1. Results using a plane as the initial water level

The results presented here were calculated with 4 GPUs. We show the domain's decomposition and its bathymetry in Figure 13, and the solutions for different times in Figure 14 .

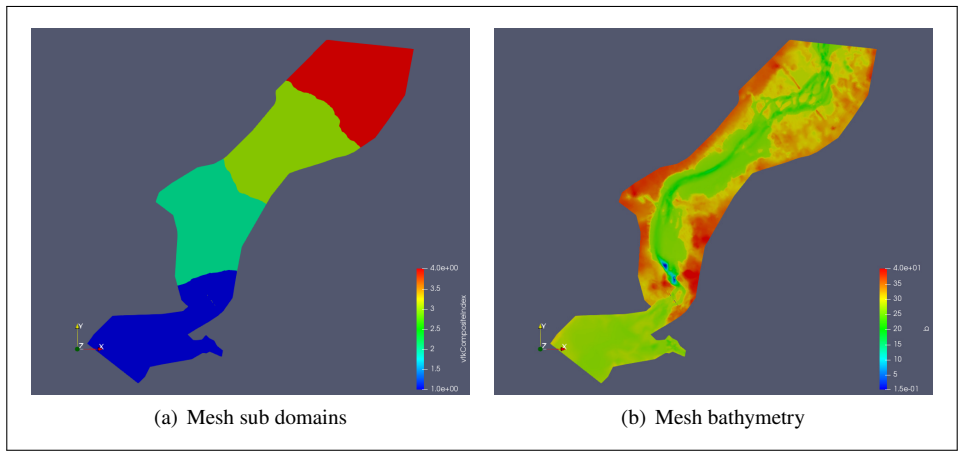

(a) Mesh sub domains          (b) Mesh bathymetry

**Figure 13:** Mille Îles river domain divided into 4 sub-domains on the left (a) and bathymetry on the right (b)

The results shown in Figure 14 were generated starting from an initialization with a plane orthogonal to the z-axis. We can see in Figure 14(a) that the entrance of the domain (in the bottom left corner) is not completely wet, and that the more the simulation advances, the more the domain fills and the more clearly we can see the river current forming.

Using 4 GPUs on the 740,000-cell mesh, we were able to calculate the 7 *h* 30 *min* of the simulation in 10 *min*. Obviously, this time is highly dependent upon the refinement of the mesh; the more the mesh is refined, the smaller the characteristic distance in the calculation of the CFL condition and the more iterations will be necessary to arrive at the final time.

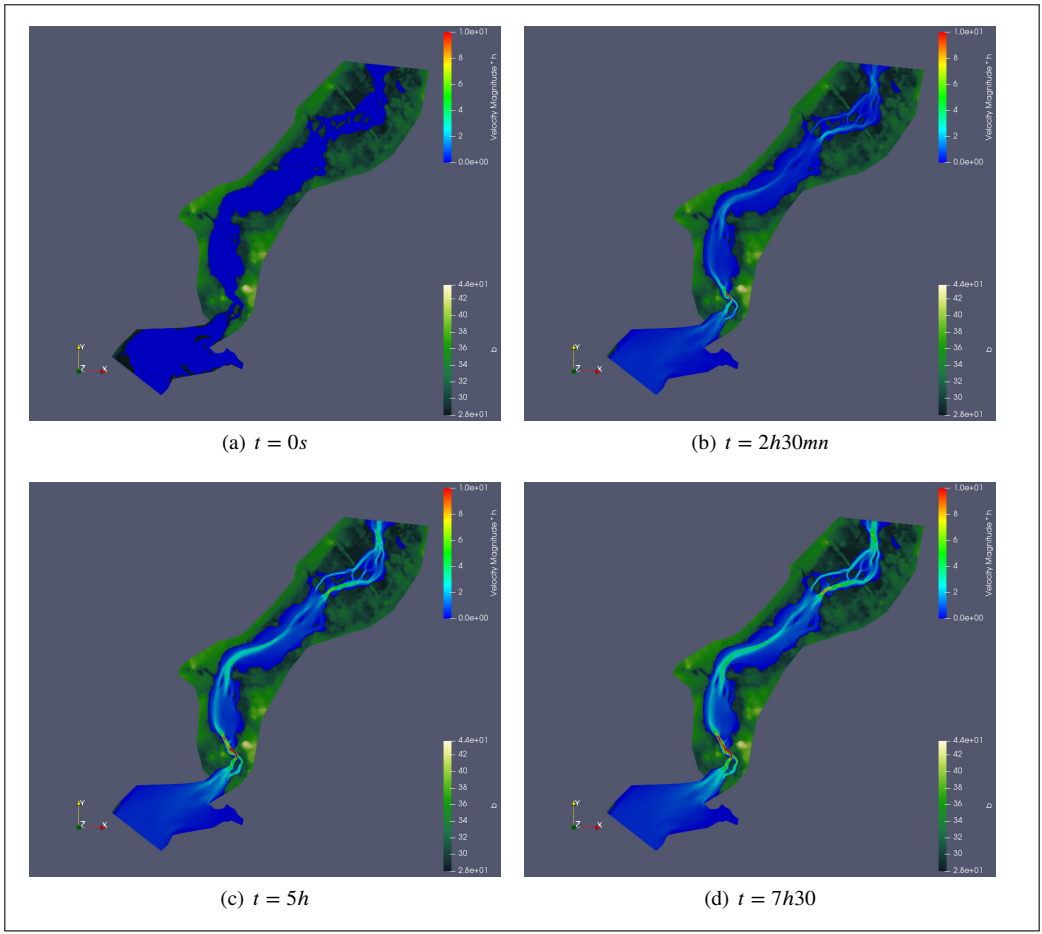

**Figure 14:** Solutions for the Mille Îles river case on a mesh of 740,000 elements using 4GPUs, dry domain colored according to bathymetry, wet domain colored according to $||h\mathbf{V}||_2$

### 6.4.2. Results of a dummy dam failure mode

Here we initialize the solution as a one-dimensional Riemann problem. On the left of the discontinuity, a water height of 30 $m$ is defined, and on the right, a height of 29 $m$. The initial velocities are zero.

We present here the solutions for different instants in order to show the usefulness of a more refined mesh. In Figure 15, the solutions on a mesh of 740,000 elements are on the left, and the solutions on a mesh of 11 million elements are on the right.

Figure 15 shows that the solutions are very close. Nevertheless, it is obvious that the solution is more finely defined starting from the initialization on the 11 million-cell mesh. We can see that the solution is finer at the end of 30 $s$, especially on the wave front. Finally, the usefulness of a fine mesh is demonstrated at the end of 120 $s$, as we can clearly observe the impact of the bridge's pillars on the watercourse, whereas with the 400,000-cell mesh the impact is practically invisible.

### 6.4.3. Visualization of the flood lines

Once we have generated a stable solution for an average flow, we can move on to phase 2 of the simulations. Figure 16 illustrates the flood lines for an inflow of 800 $m^3/s$ (in black) and an inflow of 1100 $m^3/s$ (in red) on the Mille Îles river domain.

In a real case scenario, we launch several hundred simulations by sampling the parameters, such as the inflow rate

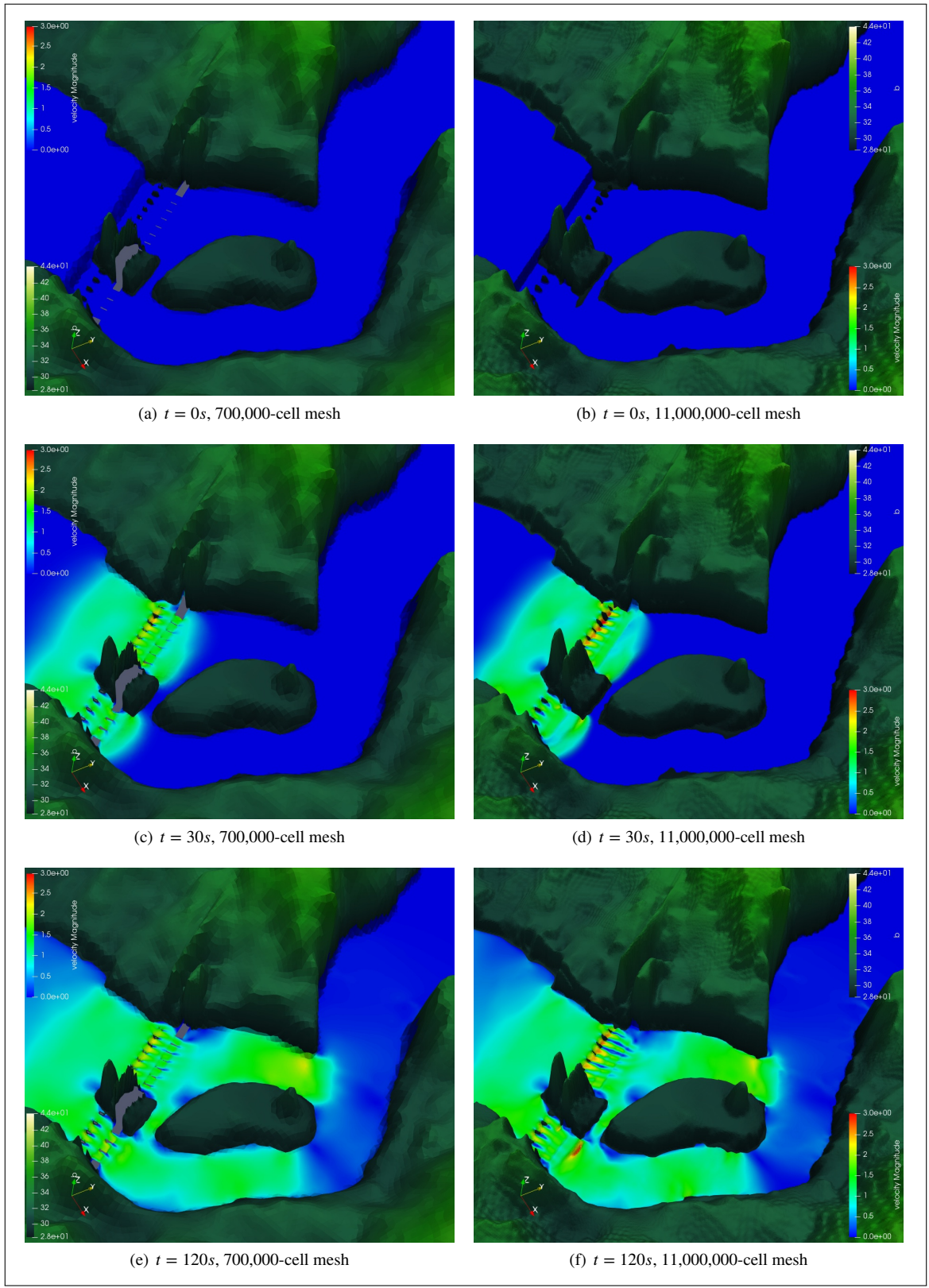

(a) $t = 0s$, 700,000-cell mesh

(b) $t = 0s$, 11,000,000-cell mesh

(c) $t = 30s$, 700,000-cell mesh

(d) $t = 30s$, 11,000,000-cell mesh

(e) $t = 120s$, 700,000-cell mesh

(f) $t = 120s$, 11,000,000-cell mesh

**Figure 15:** Solutions at different times for a problem with a fictitious dam failure on the Mille Îles river.

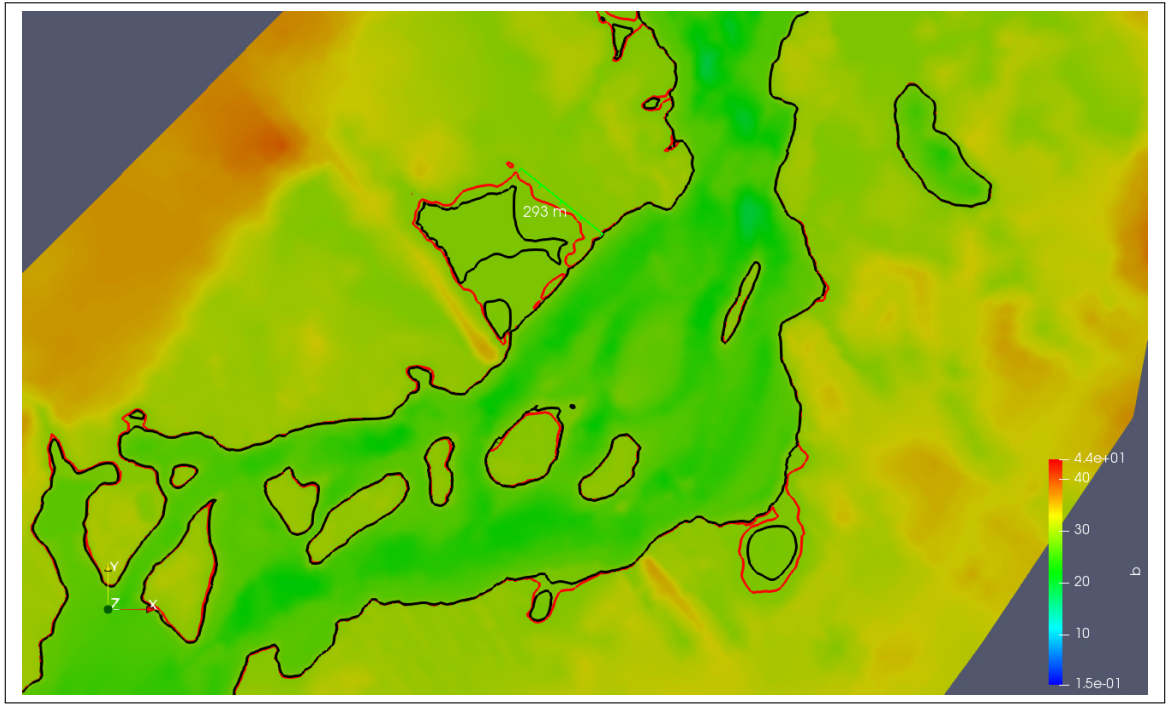

**Figure 16:** Flood lines for an inflow of 800 $m^3/s$ in black and an inflow of 1100 $m^3/s$ in red, superimposed on the bathymetry of the downstream domain of the Mille Îles river

or Manning's number. The goal is to build a database and then perform statistical studies like those in Abdedou and Soulaimani (2018). This database can also be used to train machine learning algorithms, as in Jacquier et al. (2021).

### 6.4.4. Speed-up and efficiency
Figure 17 presents the speed-up and efficiencies for two sizes of domain meshes, a version with 740,000 cells and another with 11,700,000 cells.

The results are quite identical to those of section 6.3.2. It appears that 300,000 to 500,000 elements is a good compromise with which to achieve an appropriate level of efficiency.

## 6.5. Case of the Montreal archipelago
Here we present a mesh of an area of the Montreal archipelago domain, which was provided by the *Communauté métropolitaine de Montréal* (CMM). The domain mainly includes the Mille Îles river, the Prairies River and the St. Lawrence. The domain therefore has several entrances, seven in total counting the small tributaries, and a single exit which corresponds to the St-Lawrence river at the top of the domain.

Figure 18(a) illustrates the complete domain, and Figure 18(b) shows the wet domain for a stabilized solution with a total incoming flow of 15090 $m^3/s$, an average inflow rate for this domain.

### 6.5.1. Solutions
The results presented here are calculated with four GPUs. Figure 19(a) shows the domain decomposition, divided into four sub-domains, and the bathymetry is illustrated in Figure 19(b). Next, we present the solutions for four different times ($t = 0$ $s$, 15 $mn$, 30 $mn$ and 45 $mn$) in Figure 20 .

---

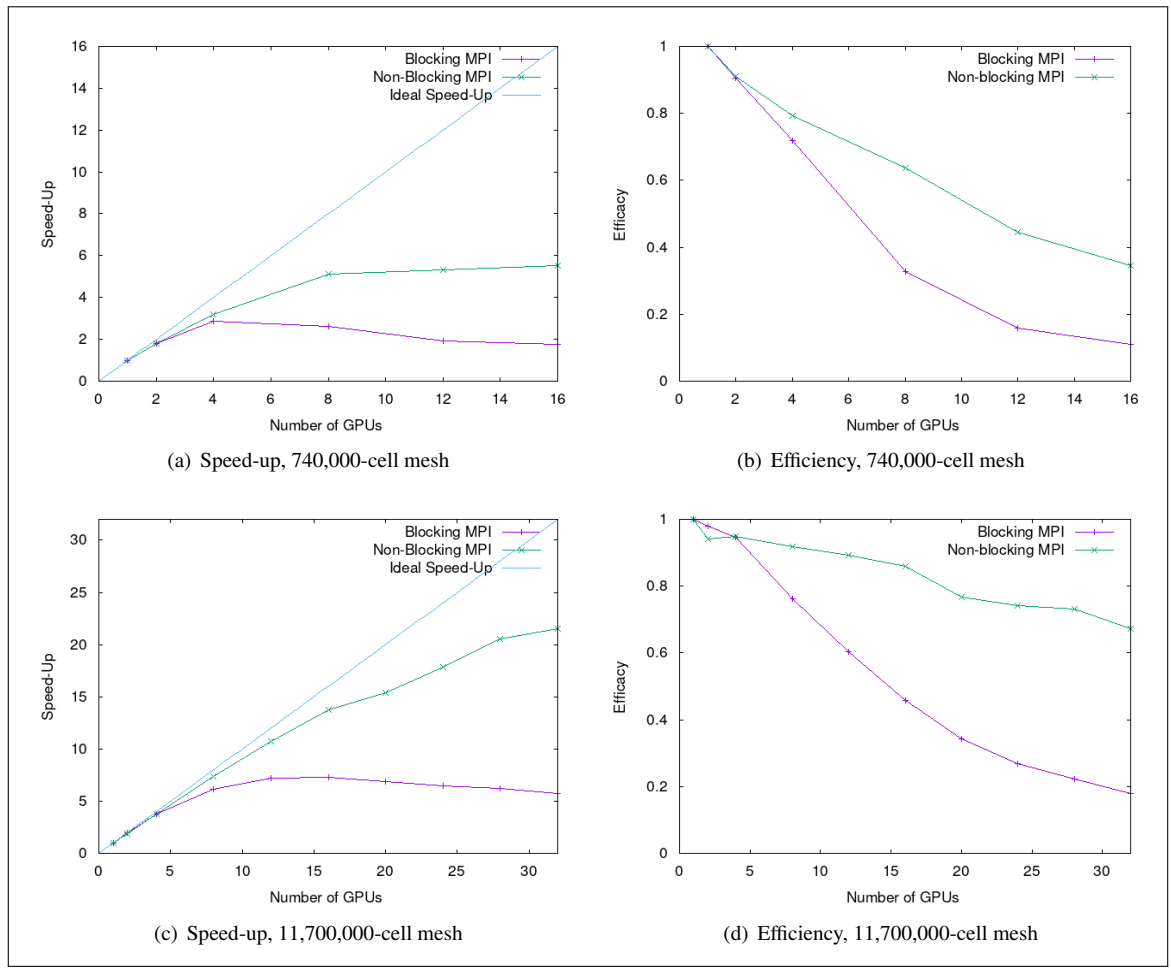

(a) Speed-up, 740,000-cell mesh

(b) Efficiency, 740,000-cell mesh

(c) Speed-up, 11,700,000-cell mesh

(d) Efficiency, 11,700,000-cell mesh

**Figure 17:** Speed-up and efficiency for multiple meshes of the Mille Îles river domain

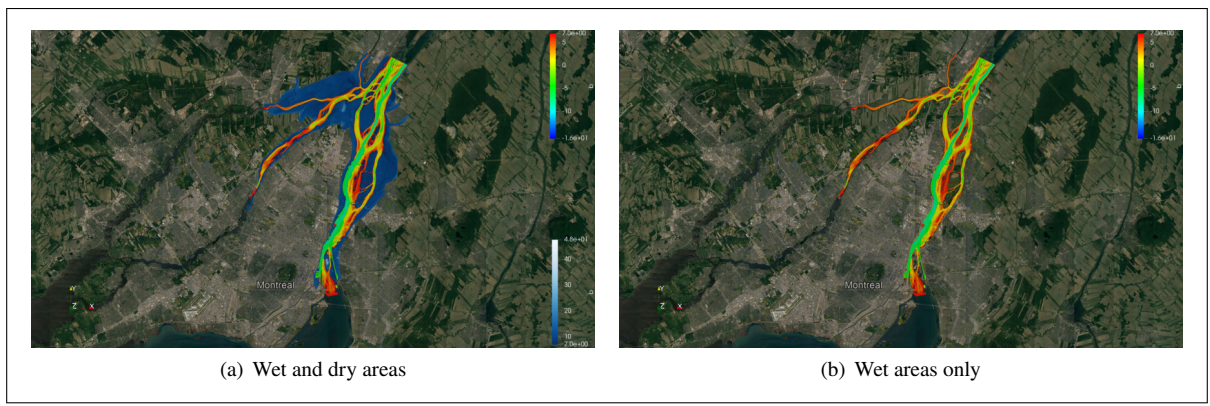

(a) Wet and dry areas

(b) Wet areas only

**Figure 18:** Area of the Montreal archipelago superimposed on a satellite image generated with Google Earth Pro. Wet and dry areas (a); wet areas only (b).

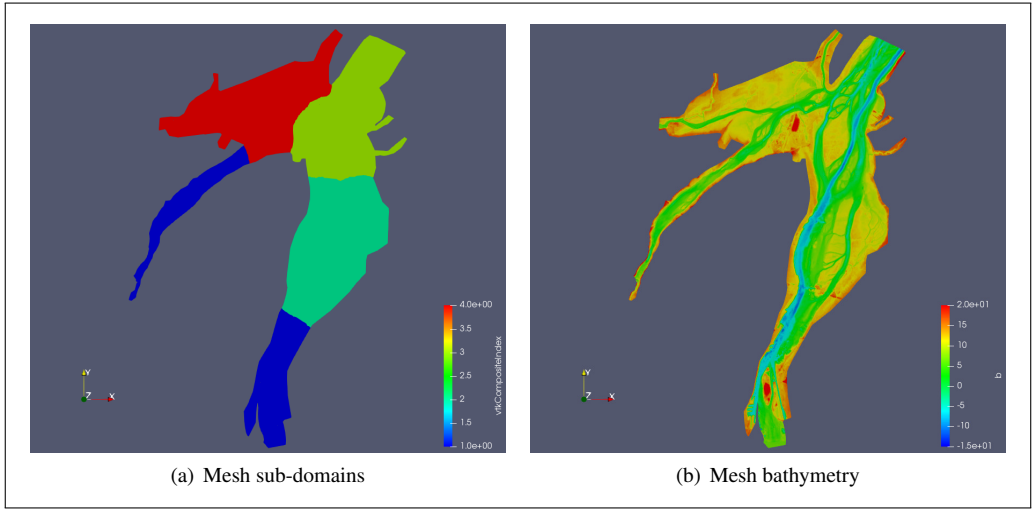

(a) Mesh sub-domains      (b) Mesh bathymetry

**Figure 19:** Domain of the Montreal archipelago divided into four sub-domains on the left (a) and its bathymetry on the right (b).

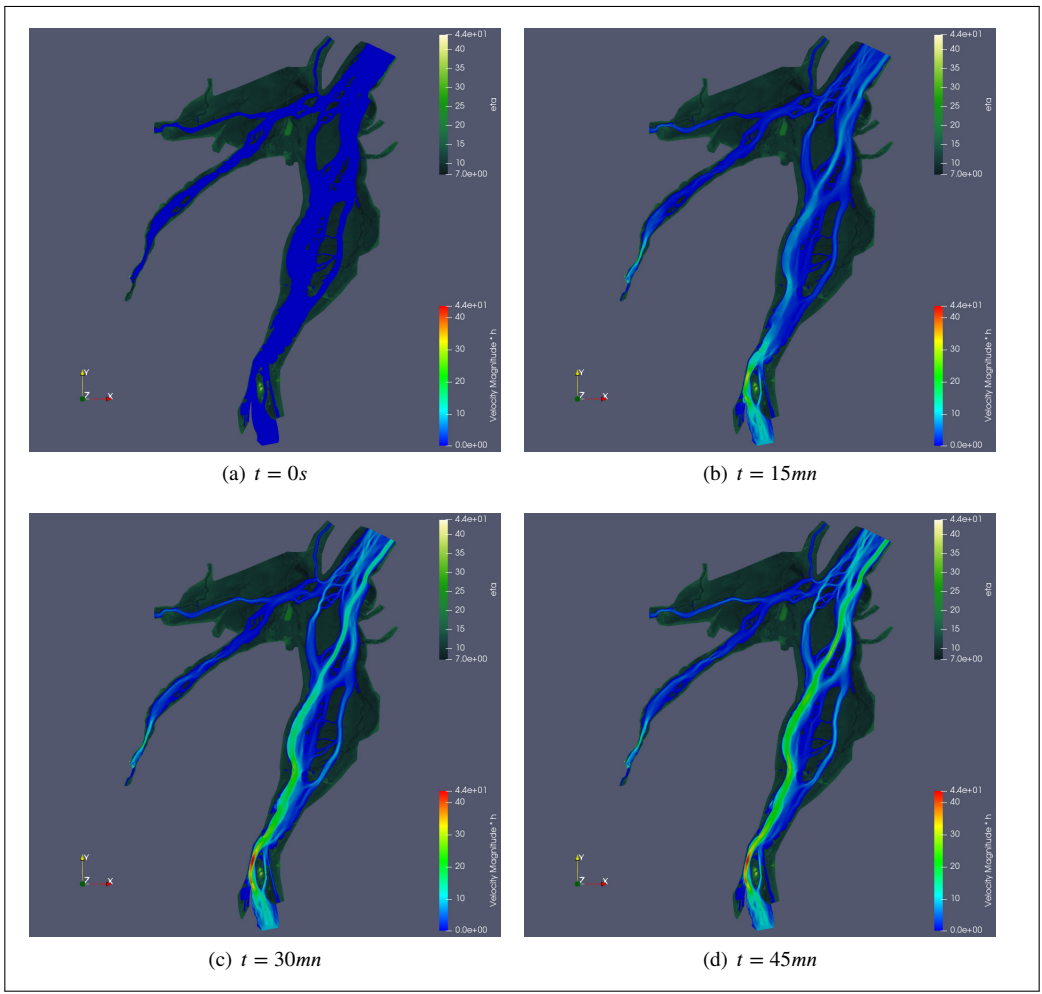

(a) $t = 0s$      (b) $t = 15mn$

(c) $t = 30mn$      (d) $t = 45mn$

**Figure 20:** Solutions at four times ($t = 0$ $s$, 15 $mn$, 30 $mn$ and 45 $mn$) for the Montreal Archipelago River on a mesh of 690,000 elements using four GPUs; dry domain colored according to bathymetry, wet domain colored according to $||h\mathbf{V}||_2$

The results shown in figure 14 were generated starting from an initialization with an inclined plane. Using an inclined plane rather than a plane orthogonal to the z-axis allows for faster initialization, as it is easier for the fluid to gain speed in the downstream direction. In figure 20, we only show the solutions for early times in the simulation; it took 11 *h* of simulation to have a steady state solution. These 11 *h* of simulation took 4 GPUs 1 hour to perform the calculations.

### 6.6. Comparison of the pure MPI CPU version to the MPI CUDA multi-GPU version

A multi-CPU version of **CuteFlow** was developed from the domain decomposition shown in 4. This version uses the same algorithm as the multi-GPU version, the only difference is that the functions written to be executed on the GPU were re-written in order to be executed on the CPU. These modifications mainly consisted of putting back the loops over the mesh cells inside the functions.

We tried using both *ifort* and *gfort* to compile the code, and achieved better results by using *ifort -O3 -xHost -ipo*, which we then used to create the results shown in figure 21. Both the multi-CPU and multi-GPU versions were launched on the same simulation using the same mesh sizes and the same parameters, while we varied the number of CPU cores and the number of GPUs, respectively, utilized to produce the results shown in Figure 21.

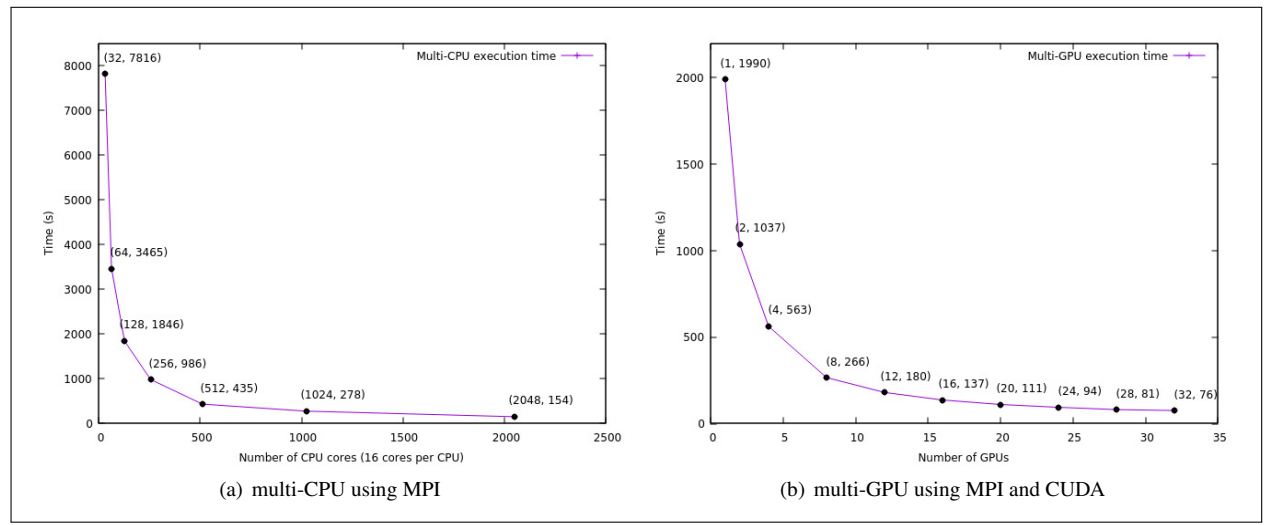

**Figure 21:** Execution times of the multi-CPU (a) and the multi-GPU (b) versions on an 11M-cell mesh

As can be seen in Figure 21, close to 96 and 128 CPU cores are needed per GPU to obtain the same multi-CPU performance as the multi-GPU version. For example, using 1024 CPU cores, the multi-CPU version takes 278 seconds to complete, whereas using 8 GPUs, it takes 266 seconds for the multi-GPU version to complete. In this case it takes 128 CPU cores per GPU to equal the performance of the multi-GPU version.

We can see that the minimum execution time is also shorter for the multi-GPU version. This is due to the use of many more sub-domains in the multi-CPU version, which leads to significantly more memory exchanges than for the multi-GPU version. This issue has been addressed in some studies by combining MPI with OpenMP, as in Shang (2014); Yilmaz et al. (2009); Jacobsen and Senocak (2013). Based on these articles, the hybrid OpenMP/MPI approach sometimes gives better performances than the pure MPI method, but we believe that in our case the multi-GPU version will always be better in terms of cost and resource utilization on computer clusters.

# 7. Conclusion

This paper has presented the different stages of parallelizing a time-explicit finite volume solver for the resolution SWE on a multi-GPU architecture. We explained the process of the SWE resolution by means of finite volume methods, with particular emphasis on the Riemann solvers used during this step.

Next, the process of porting a time-explicit finite volume solver on a multi-GPU architecture using MPI and CUDA-Fortran was detailed. The METIS library was used to tackle domain decomposition on the 2D unstructured triangular meshes of interest. During this stage, special attention must be paid to the numbering of the cells to be sent and received by each sub-domain in anticipation of the MPI memory exchanges. This section also explained the use of MPI and a CUDA-Aware version of OpenMPI in solving the SWE.

After showing the general functioning of our in-house code for the resolution of the SWE in its multi-GPU version, we presented the results for several different cases. Efficiencies of more than 80% were reported for meshes of 300,000 to 500,000 elements per GPU, no matter which mesh size was used. This means that choosing which mesh size to process only depends upon the number of GPUs available.

Finally, we compared the multi-GPU version to the pure MPI multi-CPU version and found that about 96 to 128 CPU cores are needed to equal the performance of a single GPU. The multi-GPU version allows us to obtain the same performances as using 1024 CPU cores with the multi-CPU version but with only 8 GPUs. This result clearly shows how useful and efficient multi-GPU versions can be.

We hope that the ability to use as many GPUs as needed with multi-GPU codes such as the one developed here, and their better scaling than their multi-CPU counterpart, may lead to the creation of very large river meshes of more than 100M cells.

Based on the work conducted here, we put forth the following suggestions for future studies. The MPI memory exchange needs to be further optimized by activating functionalities such as GPU-Direct RDMA on the computer clusters. This could lead to a better scaling of the CUDA MPI approach over many computer nodes. Load balancing needs to be utilized to leverage the CPU cores that are not used to control the GPUs, as done in Xu et al. (2014); Borrell et al. (2020); Fang et al. (2019). Such methods are able to make use of the idle CPU cores that are not used in the approaches proposed here, and could greatly increase the efficiency of the computer nodes. The approach proposed here should be tested on much larger meshes that have practical utility, for example, on a much more refined (10- to 100 million-cell) mesh of the Montreal archipelago.